\documentclass[useAMS,usenatbib,usegraphicx]{mn2e}
\def\healpix{H{\sc ealpix }}
\def\glesp{G{\sc lesp }}

\def\etal{et al.}

\def\alm{a_{\ell m}}
\def\Ylm{Y_{\ell m}}
\def\Cl{C_{\ell}}

\def\summ{\sum_{m=-\ell}^{\ell}}
\def\suml{\sum_{\ell=0}^{\infty}}
\def\lm{\ell, m}

\def\Cs{{\bf C}}
\def\Si{{\bf S}}

\def\l,m{\ell,m}
\def\l{\ell}

\title[Fast simulation of the CMB PMF]
{Fast simulation of the whole-sky CMB map in the presence of primordial magnetic field}
\author[Pavel Naselsky,Jaiseung Kim]{Pavel Naselsky$^{1}$\thanks{E-mail:
naselsky@nbi.dk}, Jaiseung Kim$^{1}$\\ 
$^{1}$ Niels Bohr Institute, Blegdamsvej 17, DK-2100 Copenhagen, Denmark}

\begin{document}
\date{}
\pagerange{\pageref{firstpage}--\pageref{lastpage}} \pubyear{2006}
\maketitle
\label{firstpage}
\begin{abstract}
We present a novel method for generation of sets of the Cosmic Microwave Background (CMB) anisotropy maps, which
reproduces the $\Delta \l=2$ correlations associated with Alfv\'en turbulence. The method is based on the non-linear transformation of the CMB maps, which is obtained from the Monte Carlo simulation of the statistically isotropic Gaussian signal. Our method is computationally fast and efficient.

We have applied two estimators (the cross-correlation estimator in multipole domain for $\l+1,m$ and $\l-1,m$ modes and circular phase moments) to test the statistical properties of derived maps. Both of these statistics confirm the effectiveness of our generation method. We believe that our method can be useful for fast generation of the non-Gaussian maps in the presence of the primordial magnetic field, and be a valuable tool for the non-Gaussianity investigation of the CMB in the framework of the future PLANCK data analysis. 
\end{abstract}

\begin{keywords}
magnetic fields  -- cosmology: cosmic microwave background -- methods: data analysis
\end{keywords}

\section{Introduction}

The primordial magnetic field (PMF) could be one of the most impressive non-Gaussian relic of the epoch of inflation (see Branderburg, Enqviest and Olesen (1996), Durrer, Kahniashvili and Yates (1998), Mack, Kahniashvili and Kosowsky (2002), Subramanian, Seshadri and Barrow (2003), Chen et al. (2004), Naselsky et al. (2004), Giovannini (2007), and etc.) which can be tested by observation of CMB anisotropy and polarization.    

The presence of the PMF with present strength $\sim 1-10$ nG and coherent scale comparable or above the present day horizon can induce and support vorticity  or Alfv\'en turbulence 
 before and during the epoch of hydrogen recombination and via interaction with the CMB photons can produce non-adiabatic tail of the CMB anisotropy and polarization. The existence of a large-scale coherence of the PMF is a potential source of the preferred angular directions in the CMB map, which manifest themself as a sources of statistical anisotropy and non-Gaussianity of the CMB. Complimentary to the models of non-trivial topology such as Bianchi $VII_h$, cosmological strings and etc., the PMF can significantly contribute to the low multipole part of the CMB power spectrum, causing some morphological peculiarities of the CMB sky. 

 As shown in  Chen et al, 2004, the Alfv\'en turbulence arising from primordial magnetic field (statistically homogeneous and isotropic at very large scales), should have non-Gaussian properties, because of quadratic dependence of the vorticity amplitude on the magnetic strength B. In the reference coordinate system of $\textbf{z}||{\bf B}$, magnetic field induces correlation between the $a_{l-1,m}$ and $a_{l+1,m}$ harmonic coefficients of the CMB temperature anisotropy \footnote{ 
In general coordinate systems where the orientation of the magnetic vector \textbf{B} is arbitrary, 
there exist another correlations between $a_{l,m}$ and $a_{l',m'}$.}. Hereafter, our choice of the reference coordinate system ($\textbf{z}||{\bf B}$) is implicit through this paper.
 
Since the PMF has drawn very serious attention after the WMAP experiment, the investigation of possible observational traces from the PMF has been included in the data analysis agenda for the PLANCK surveyor mission. 
In the framework of the non-Gaussianity investigation of the CMB observation by the PLANCK we need to generate about $10^3-10^5$ realizations of the CMB anisotropy maps in the presence of the PMF in order to test their statistical properties by various methods. 
Keeping ``the  open window'' for other sophisticated methods, in this paper we propose a fast generation method of the CMB PMF map, based on non-linear transformation of the statistically homogeneous and isotropic Gaussian Random Field into non-Gaussian and anisotropic one.  

 Let us outline the basic requirements for any methods of modeling the CMB PMF. 
Firstly, the method should correctly reproduce the statistical properties of the CMB PMF signal, namely, the power spectrum and the correlations between $a_{l-1,m}$ and $a_{l+1,m}$ harmonic coefficients. Secondly, the method should be computationally simple  and fast prefably at the level of Random Gaussian Field (RGF) simulation. Thirdly, it should possess good scalability with increasing angular resolution of the maps. For instance, the method should cover the multipole range up to $l\sim 4-6\times 10^3$ for the PLANCK mission. 
The outline of this paper is as follows. In Section 2 we briefly discuss the basic properties of the CMB PMF. Section 3 is devoted to description of the non-linear generator (hereafter PMF generator) and discusses the properties of derived signal. In Section 4 we present the analysis of correlation between different multipole moments and show that our PMF generator reproduce correct theoretical properties of the CMB PMF signal. In Section 5 we make the phase analysis on the generated CMB PMF maps in order to show that morphology of generated maps is correctly reproduced by our PMF generator. In conclusion we summarize the results of investigations, and mark problems relavant to the future PLANCK data analysis.

 \section{Statistics of the CMB anisotropy generated by the Alfv\'en turbulence}
One of the method to characterize the temperature anisotropy in the direction $\vec{n}=(\theta,\varphi)$ is decomposition of the signal $\Delta T(\vec{n})$ in spherical
 spherical harmonics:
\begin{equation}
\Delta T(\theta,\varphi)=\suml \summ |\alm|e^{i\phi_{lm}} \Ylm (\theta,\varphi),
\label{eq1}
\end{equation}
where $|\alm|$ and $\phi_{lm}$ are the modulus and phase of the expansion coefficients respectively.

Homogeneous and isotropic CMB Gaussian random fields (GRFs) possess $\ell,m$ modes whose real
and imaginary parts are Gaussian and mutually independent. The statistical
properties  of the RGF are then completely specified by its angular power spectrum
$\Cl^{cmb}$,  
\begin{equation}
\langle  a^{cmb}_{\ell^{ } m^{ }} (a^{cmb})^{*}_{\ell^{'} m^{'}}
\rangle = \Cl^{cmb} \; \delta_{\ell^{ } \ell^{'}} \delta_{m^{} m^{'}}.  
\label{eq2}
\end{equation}
In other words, from the Central Limit Theorem their phases
\begin{equation}
\phi^{cmb}_{\ell m}=\tan^{-1}\frac{\Im (\alm^{cmb})}{\Re (\alm^{cmb})}
\label{eq3}
\end{equation}
are randomly and uniformly distributed in the range $[0,2\pi]$, and the amplitude $|\alm|$ follows Ryley distribution. We hereafter denote the pure Gaussian CMB signal by the superscript ``{\it G}''.
The bracket in Eq. \ref{eq2} denotes average over an ensemble of realizations. The expectational value of the power spectrum
over an ensemble corresponds to theoretical predictions. For the magnetized Universe, the helical part of vorticity does not contribute to the power spectrum, but induces off-diagonal correlations between $\ell-1,m$ and $\ell+1,m$ harmonic coefficients:
\begin{eqnarray}
D_l(m)&=&\langle  a^{cmb}_{\ell-1, m} (a^{cmb})^{*}_{\ell^{'}+1, m}\rangle \nonumber\\
&=&\langle  (a^{cmb})^{*}_{\ell+1, m^{ }} (a^{cmb})_{\ell^{'}-1, m^{}}\rangle\label{eq4}
\end{eqnarray}
Moreover, since the statistical isotropy of the CMB has been broken by the presence of the PMF, the power spectrum 
\begin{eqnarray}
 C_{\ell}(m)=\langle  a^{cmb}_{\ell^{ } m^{ }} (a^{cmb})^{*}_{\ell m}
\rangle
\label{eq5}
\end{eqnarray}
now reflect directly this statistical anisotropy through the dependence of $C_{\ell}$ on $m$.

In the paper Durrer, Kahniashvili and Yates (1998) (hereafter DKY) it was shown that  the power spectrum $C_{\ell}(m)$ and auto-correlator  
$D_{\ell}(m) $ are given by
\begin{eqnarray}
\nonumber \\
C_{\ell}(m)\propto \frac{2^{n+1}\Gamma(-n-1)\Gamma(l+\frac{n}{2}+\frac{3}{2})}
{\Gamma(-\frac{n}{2})^2\Gamma(l-\frac{n}{2}+\frac{1}{2})}\times \nonumber \\
\frac{2l^4 +4l^3 -l^2 -3l +(6 -2l -2l^2)m^2}{(2l-1)(2l+3)};\nonumber \\
D_{\ell}(m)\propto\frac{2^{n+2}}{|n+1|}\frac{(\Gamma(-n-1)\Gamma(l+\frac{n}{2}+\frac{3}{2})}{\Gamma(-\frac{n}{2})^2\Gamma(l-\frac{n}{2}+\frac{1}{2})}\times  \nonumber \\
(l-1)(l+2)\left(\frac{(l+m+1)(l-m+1)(l+m)(l-m)}{(2l-1)(2l+1)^2(2l+3)}\right)^{\frac{1}{2}},\nonumber \\
\label{eq6}
\end{eqnarray}
where $n$ is the power index of the Alfv\'en turbulence power spectrum and Eq. \ref{eq6} is valid in the range $-7< n<-1$.
 The simplest way to detect the presence of a homogeneous magnetic field in the Universe is based on the arithmetic means over $m$ of the two spectra, proposed in DKY
\begin{eqnarray}
 \overline C_{\ell}&\equiv&\frac{1}{2\ell+1}\sum_{m=-\l}^{\l}\langle a^{*}_{\lm}a^{}_{\lm}\rangle,\nonumber\\
 \overline D_{\ell}&\equiv&\frac{1}{2\ell+1}\sum_{m=-\l}^{\l}\langle a^{*}_{\ell-1,m}a^{}_{\ell+1,m}\rangle.
\label{eq7}
\end{eqnarray}

DKY find that
\begin{eqnarray}
\overline C_{\ell}&\simeq& A_0\left(\frac{t_{dec}}{t_0}\right)(k_0 t_0)^{-(n+3)},\nonumber\\
&\times& v^2_A\frac{2^{n+1}\Gamma(-n-1)}{3\Gamma^2(-n/2)}\l^{n+3},\;n<-1\nonumber\\
\overline C_{\ell}/\overline  D_{\ell}&=&|n+1|\left[\frac{\Gamma(-\frac{n+1}{2})}{\Gamma(-\frac{n}{2})}\right]^2,\; n<-1\nonumber\\
 \overline C_\l\simeq \overline D_\l&\simeq&\frac{v^2_A A_0}{2\pi(k_0t_0)^2}\left(\frac{t_{dec}}{t_0}\right)^2 \l^2,\;\;n>-1.
\label{eq8}
\end{eqnarray}
where $A_0$ is a dimensionless normalization constant, $t_{dec}/t_0$ is the decoupling time to the present time ratio, and
$k_0$ is the damping wave number. Though it is not shown explicitly in Eq. \ref{eq8}, the values of $D_\l(m)$ may also be negative.
As one can see from Eq.(\ref{eq8}), the statistical isotropy of the CMB signal from the PMF is now restored, since the power spectrum $\overline C_{\ell}$ does not depend on the $m$. Thus, if we model the statistical properties of the CMB PMF by using $\overline C_{\ell}$ and $\overline D_{\ell}$ powers as (Chen et al, 2004), we are actually dealing with statistically isotropic, but non-Gaussian random field. This model is very useful for the situation, where the coherence scale of the PMF is much smaller than the present-day horizon, and the patterns of magnetic field are randomly oriented in space just as magnetic domains in ferromagnetic materials. 

Using the limiting values for $A_0$ and $k_0$ in Eq. \ref{eq8}, the followings $\overline C_{\ell}$ are obtained (Chen et al. 2004):
\begin{eqnarray}
\overline C_{\ell}&=&9.04\times 10^{-4} \l^{-2}\left(\frac{B}{10^{-9} \mathrm G}\right)^4 \mu \mathrm{K}^2,\;\; n=-5 \nonumber\\
\overline C_{\ell}&=&8.61\times 10^{2} \l^{-4}\left(\frac{B}{10^{-9} \mathrm  G}\right)^4 \mu \mathrm{K}^2,\;\; n=-7, \label{Cl_constraint}
\end{eqnarray}
where $B$ is the strength of magnetic field in the unit of Gauss.
These two cases correspond to a Harrison-Peebles-Yu-Zel'dovich scale-invariant spectrum and a possible inflation model of primordial vorticity field respectively.
The \textit{WMAP} data impose the follwing upper limit on the strength of the primordial magnetic field at 3 $\sigma$ confidence level (Chen et al. 2004):
\begin{eqnarray}
B<15\times 10^{-9}\mathrm G\;\;\;\;\;(n=-5),\nonumber\\
B<1.7\times 10^{-9} \mathrm G\;\;\;\;(n=-7).\label{B_limit} 
\end{eqnarray}

\section{Non-linear generator of the CMB PMF.}

\begin{figure}
\centering\includegraphics[scale=.27]{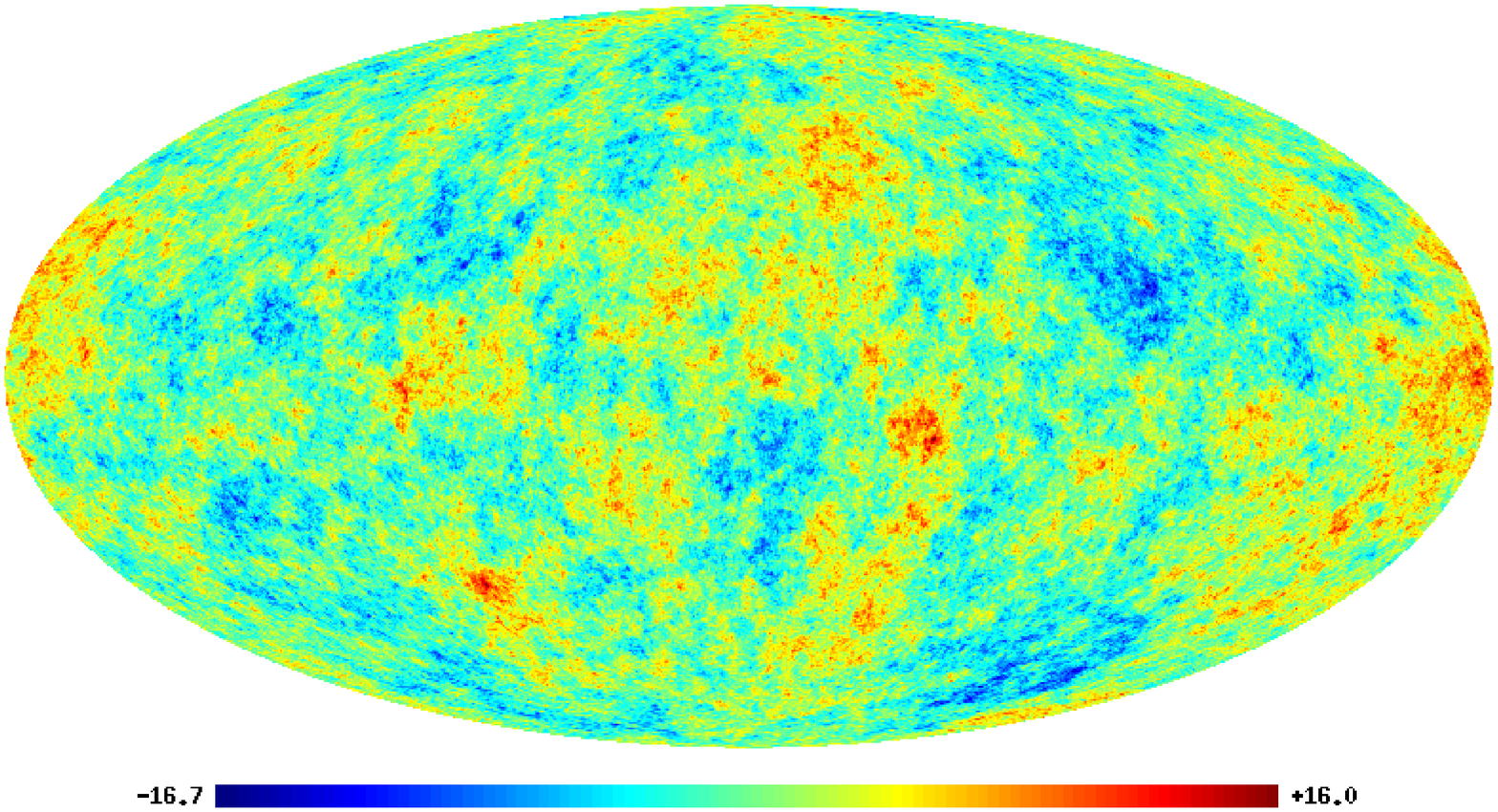}
\centering\includegraphics[scale=.27]{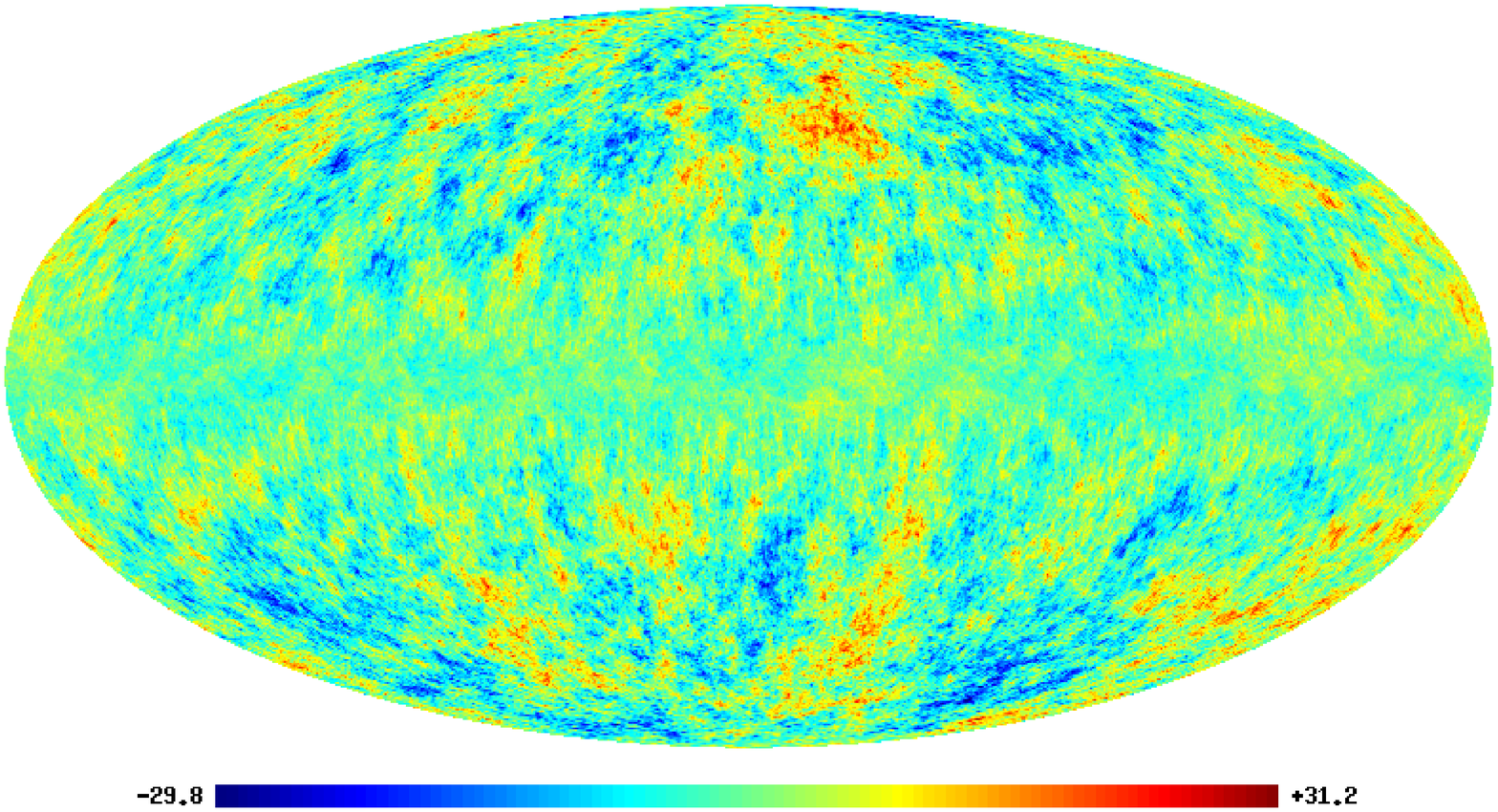}
\centering\includegraphics[scale=.27]{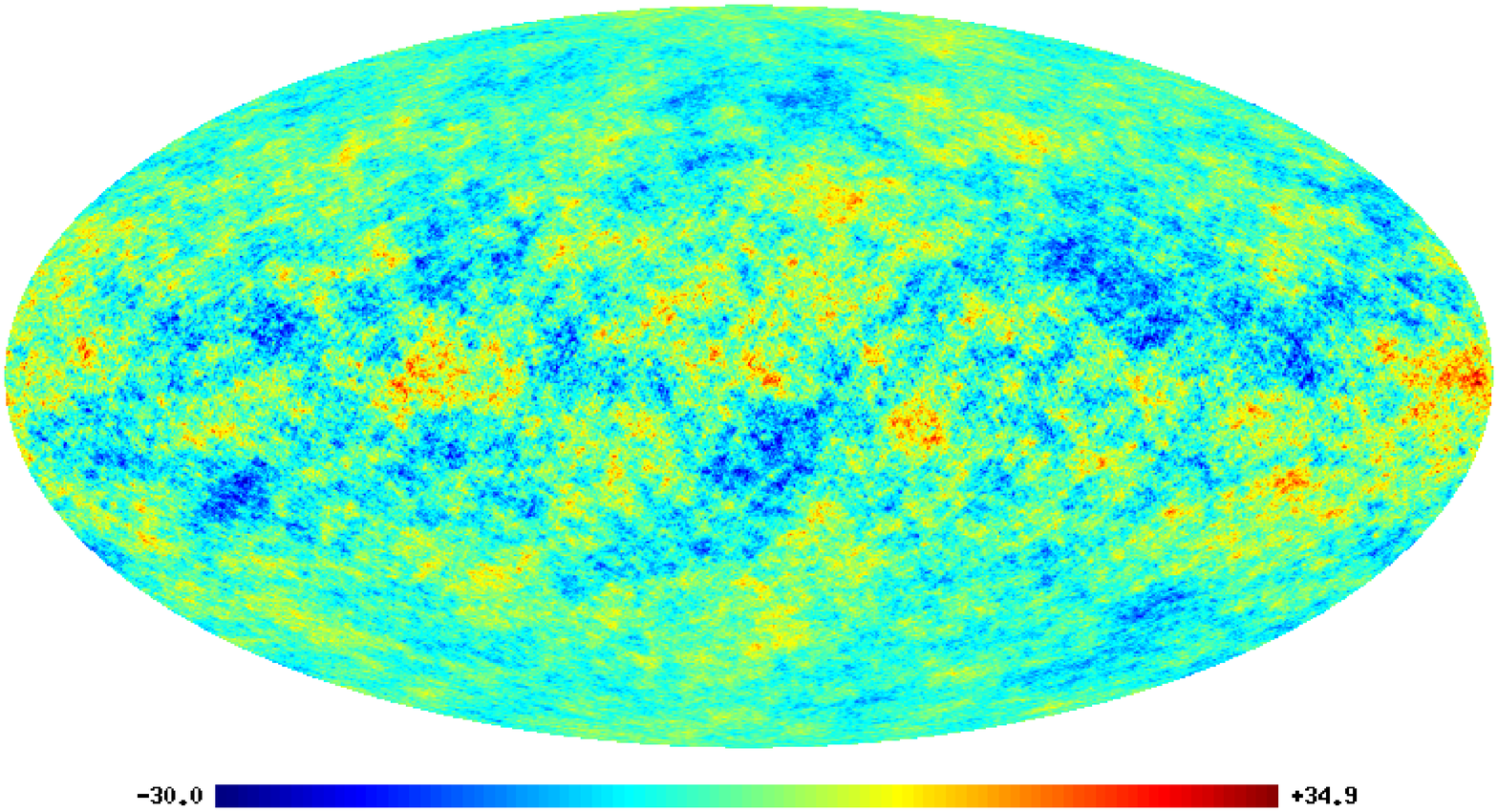}
\caption{The input map for the RGF CMB with $n=-5$ power spectrum (top). Middle - the CMB PMF for 
$\alpha = 1.66$. Bottom- the CMB PMF for $\alpha = -1.66$.}
\label{f1} 
\end{figure}

\begin{figure}
\centering\includegraphics[scale=.27]{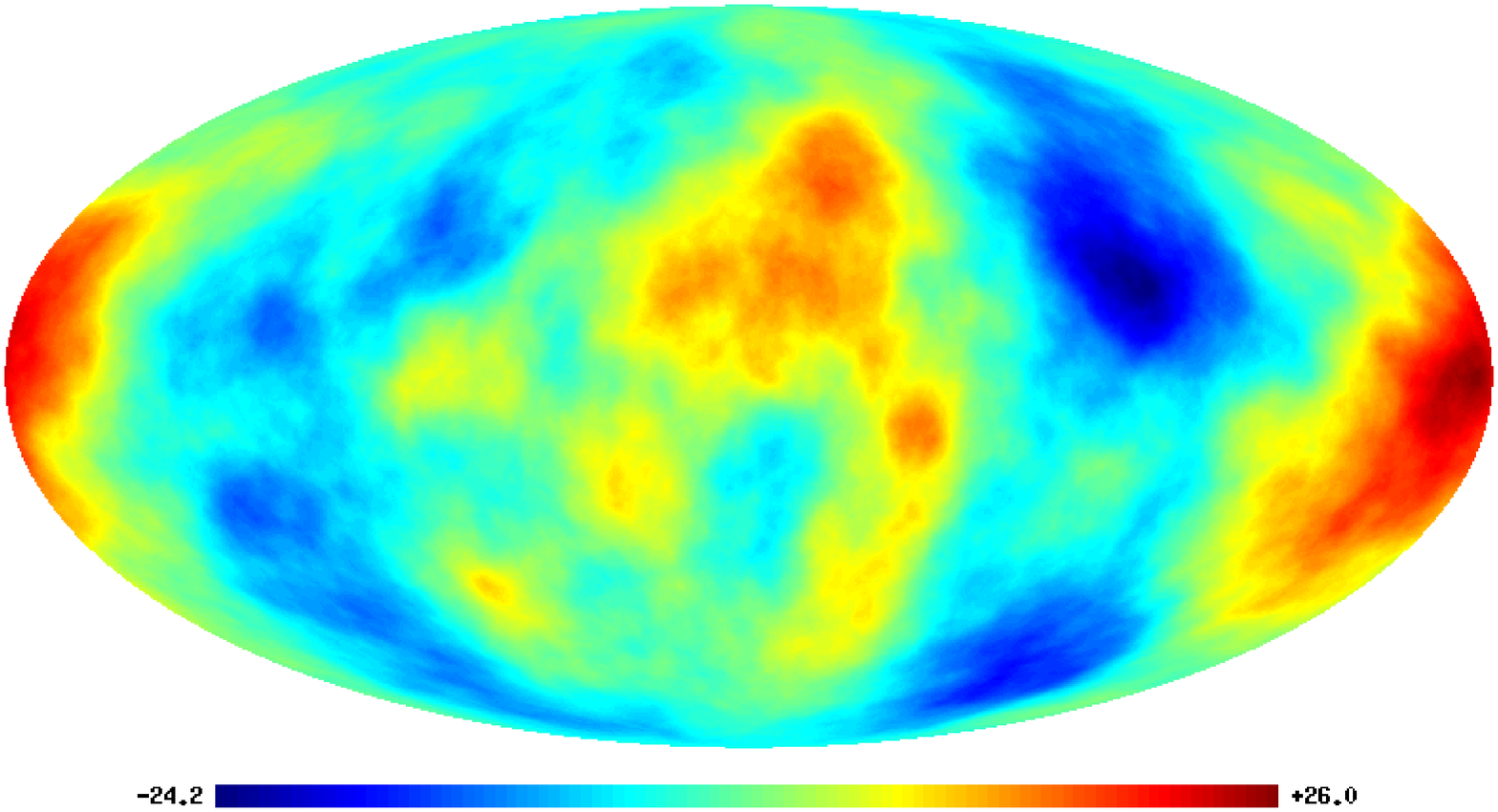}
\centering\includegraphics[scale=.27]{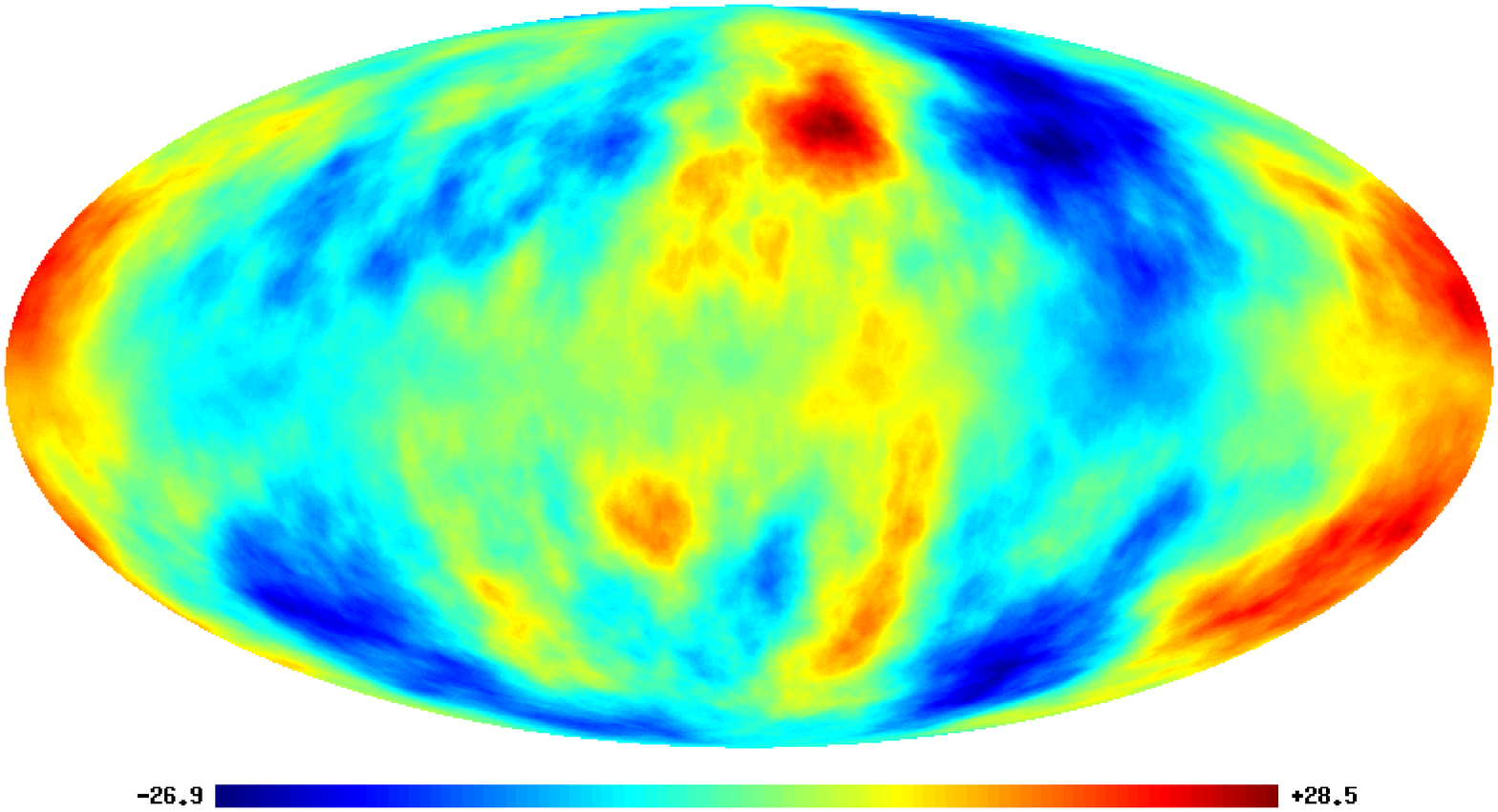}
\centering\includegraphics[scale=.27]{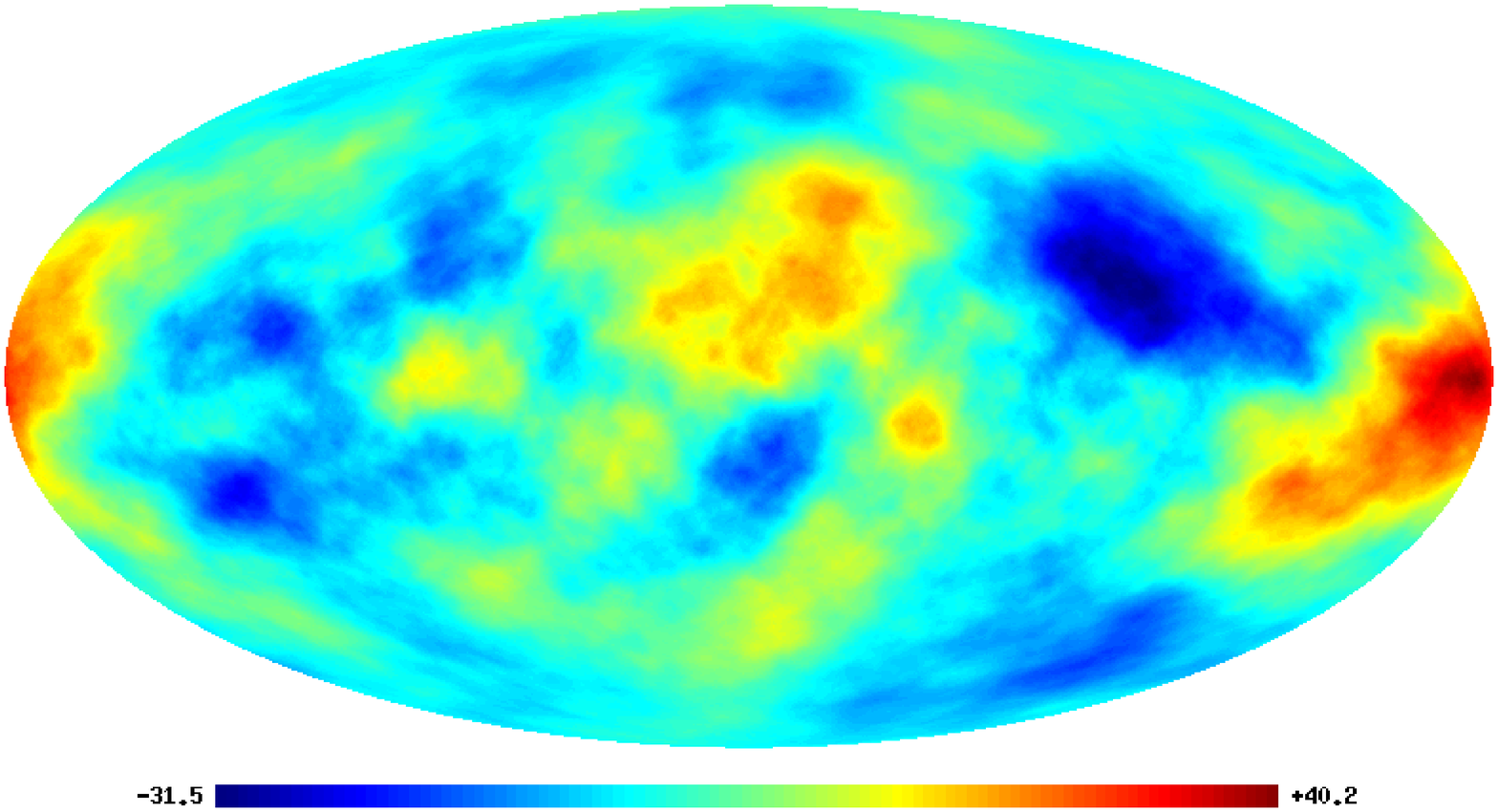}
\caption{The input map for the RGF CMB with $n=-7$ power spectrum (top). Middle - the CMB PMF for 
$\alpha= 1.51$. Bottom- the CMB PMF for $ \alpha= -1.51$.}
\label{f2} 
\end{figure}

In this section we discuss the simplest way to model the properties of the CMB PMF, which uses RGF CMB signals obtained from the Monte Carlo method. Let's start from the model, which can reproduce the average power spectra $\overline C_{\l}$ of the CMB PMF signal. For this, we generalize the method, proposed in Naselsky et al. (2006) and define a CMB map with the following harmonic coefficients:,
\begin{eqnarray} 
 c_{\lm}=a_{\lm}+\alpha\frac{|a_{\lm}|}{|a_{\l-2,m}|}a_{\l-2,m}=\nonumber\\
=a_{\lm}\left[1+\alpha\exp\left(i(\phi_{\l-2,m}-\phi_{\lm})\right)\right]
\label{eq10}
\end{eqnarray}
Note that in Eq. (\ref{eq10}) the $\alpha$-parameter is a function of the power index $n$ only, which can be easily shown by Eq. \ref{eq8} and \ref{eq16}. We draw $a_{\lm}$ in Eq. (\ref{eq10}) randonly from Gaussian distribution of the variance $C_G(\l)$. From Eq. (\ref{eq10}) one can get
\begin{eqnarray}
|c_{\lm}|^2= (1+\alpha^2)|a_{\lm}|^2+\alpha\frac{|a^{}_{\lm}|}{|a_{\l-2,m}| }\times\nonumber\\
\times\left(a^{*}_{\lm}a_{\l-2,m}+a^{}_{\lm}a^{*}_{\l-2,m}\right).\nonumber\\
\label{eq11}
\end{eqnarray}
Since $a_{\lm}$ corresponds to RGF, average over realizations and $m$ gives us
\begin{eqnarray}
\overline C_{\l}=(1+\alpha^2)\,C_G(\l).\label{eq12}
\end{eqnarray}
Thus, the power of the RGF $C_G(\l)$ is equivalent to the power of the CMB PMF, up to the factor $1+\alpha^2$.
Let's draw our attention to the following combination of the $c_{\lm}$ coefficients:
\begin{eqnarray}
c_{\l-1,m}c^{*}_{\l+1,m}+c^{*}_{\l-1,m}c^{}_{\l+1,m}=\nonumber\\
a^{*}_{\l-1,m}a^{}_{\l+1,m}+a^{}_{\l-1,m}a^{*}_{\l+1,m}+2\alpha|a^{}_{\l-1,m}||a^{}_{\l+1,m}|\nonumber\\
+\alpha\frac{|a^{}_{\l-1,m}|}{|a^{}_{\l-3,m}|}(a^{*}_{\l+1,m}a^{}_{\l-3,m}+a^{}_{\l+1,m}a^{*}_{\l-3,m})\nonumber\\
+\alpha^2\frac{|a^{}_{\l+1,m}|}{|a^{}_{\l-3,m}|}(a^{*}_{\l-1,m}a^{}_{\l-3,m}+a^{}_{\l-1,m}a^{*}_{\l-3,m}),\nonumber\\
\label{eq13}
\end{eqnarray}
Since the average over realizations gives us
\begin{eqnarray}
 D_{\l}(m)=\alpha\langle |a^{}_{\l-1,m}||a^{}_{\l+1,m}|\rangle\neq 0,
\label{eq14}
\end{eqnarray}
by averaging over $m$ we finally obtain: 
\begin{eqnarray}
\overline D_{\l}\simeq \alpha C_G(\l).\label{eq15}
\end{eqnarray}
We introduce $\gamma$ to denote the $\overline C_l$ to $\overline D_l$ ratio:
\[\gamma=\frac{\overline C_l}{\overline D_l}.\]
Using Eq. \ref{eq8}, \ref{eq12} and \ref{eq15}, we obtain the following equation for the $\alpha$-parameter:
\begin{eqnarray}
1-\gamma \alpha+\alpha^2=0, \label{alpha_equation}
\end{eqnarray}
and corresponding solutions:
\begin{eqnarray}
 \alpha=\frac{\gamma}{2}\pm \sqrt{\frac{\gamma^2}{4}-1}.\label{eq16}
\end{eqnarray}
In Table \ref{gamma}, we show the values of $\gamma$ and $\alpha$ for various $n$.
\begin{table}
\centering\caption{the value of $\alpha$ for various $n$}
\label{gamma}
\begin{tabular}{ccc}
\hline\hline 
$n$ &$\gamma$ & $\alpha$\\
\hline
-4 &$\pm$2.36 & $\pm(1.18\pm0.062)$\\
-5 &$\pm$2.26 & $\pm(1.13\pm0.53)$\\
-6 &$\pm$2.21 & $\pm(1.1\pm0.47)$\\
-7 &$\pm$2.17 & $\pm(1.09\pm0.42)$\\
\hline 
\end{tabular}
\end{table}
The generation procedure is quite straightforward.
We draw the set of $a_{\l,m}$ from Gaussian distribution of the variance $\overline C_{\l}/(1+\alpha^2)$. 
Then, we obtain $c_{\lm}$ by applying the non-linear generator in Eq. \ref{eq10} to the set of the RGF $a_{\l,m}$.
We present illustrative cases in Fig. \ref{f1} and \ref{f2}, which are obtained up to multipoles $l\le 500$, using the upper limit values of B in Eq. \ref{B_limit}. 
The top panel in Fig. \ref{f1} shows the RGF CMB map drawn from Gaussian distribution of the power spectrum $C(\l)\propto \l^{-2}$. The middle panel shows the map we have obtained, using Eq. \ref{eq10}  with $n=-5$ and the maximal positive root $\alpha= 1.66$ from Eq. (\ref{eq16}). The bottom plot is the same as the middle one except for the sign of $\alpha= -1.66$.
These figures clearly show that our PMF-generator significantly changes the morphology of the input RGF map, inducing significant $\Delta\l=2$ correlations.
Fig. \ref{f2} illustrates the properties of PMF-generator for the model with $C(\l)\propto \l^{-4}$, and $n=-7$. It should be noted that for this figure we used the same RGF CMB with that of Fig.\ref{f1} and simply rescaled the power by implementation of the filter $P(\l,n=-7,n=-5)=\overline C(\l,n=-7)/\overline C(\l,n=-5)$ (see Novikov et al, 2001). It is not surprising that for the $n=-7$ model the low multipole tail of the CMB PMF image is now dominant over high multipoles and the image has significant large angular scale modulation. From Fig.\ref{f1}-\ref{f2} one can see common feature of our PMF generation method. For a positive $\alpha$ the zone $\theta=\pi/2$ contains some stripes, while zones around polar cups retain much of morphology similar to the input map. For a negative $\alpha$ we have opposite tendency - the zones around the  polar cups are modified most. However, to characterize these distortion quantitatively we need to use appropriate estimators of non-Gaussianity and statistical anisotropy. Below we use the simplest two estimators, which will directly reflect the coupling between $\Delta\l =2$ multipoles. 

\section{Statistical properties of the CMB PMF}
\subsection{Cross-correlation coefficient between multipole momentum.}
\begin{figure}
\centering\includegraphics[scale=.43]{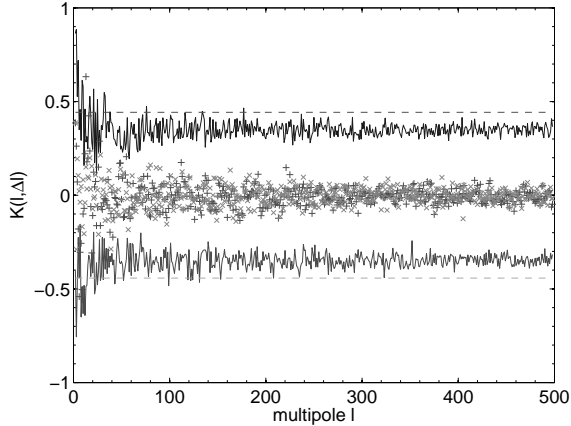}
\caption{The coefficient of cross-correlation $K(l,\Delta\l)$. The thick solid lines correspond to the
$K(l,\Delta\l=2)$ with $n=-5,\,\alpha=1.66$ (the upper curve) and $\alpha=-1.66$ (the lower curve). 
The cross and plus signs correspond to $K(l,\Delta\l=1)$ with $\alpha=\pm 1.66$ respectively.
The dashed lines correspond to $\pm \overline D_{\l}/\overline C_{\l}$.}
\label{f3} 
\end{figure}

To characterize the coupling between $\l,m$ and $\l+2,m$ modes of the CMB PMF we propose to use the coefficient of cross-correlation, defined in multipole domain as
\begin{eqnarray}
 K(l,\Delta\l)=\frac{\sum_{m=-\l}^{\l}\left(c_{\lm}c^{*}_{\l+\Delta \l,m}+c^{*}_{\lm}c_{\l+\Delta\l,m}\right)}{
2\left[\sum_{m=-\l}^{\l}|c_{\lm}|^2\sum_{m'=-\l}^{\l}|c_{\l+\Delta\l,m'}|^2\right]^{\frac{1}{2}}}\nonumber\\
\label{eq17}
\end{eqnarray}
This coefficient $K(l,\Delta\l)$ is directly related to the cross-correlation between any multipole moment $\l$ and $\l+\Delta\l$, and can be used for the analysis of $\Delta\l=2$ correlation of the CMB PMF. On the other hand, it should be noted that the $K(l,\Delta\l)$-estimator for uncorrelated signal may possess small but non-zero values, since the $K(l,\Delta\l)$-estimator is applied to a single realization of the CMB PMF. However, we expect to find the cross-correlation of $\Delta\l=2$ unusually high in comparison with correlations of $\Delta\l=1,3,5...$. Since the nominator and the denominator in Eq.(\ref{eq17}) are similar to $\overline D_{\l}$ and $\overline C_{\l}$ when averaged over realizations, the values of $K(l,\Delta\l=2)$-estimator is expected to be $\sim 1/\gamma$.
However, for each single realization we should have statistical deviation of $K(l,\Delta\l=2)$ from the expectation value. 
Nevertheless, it is important to note that the $K(l,\Delta\l=2)$-coefficient is a good measure of 
$\Delta\l=2$ cross-correlations. In Fig. \ref{f3}, significant $\Delta\l=2$ correlations of the simulated CMB PMF are clearly shown in contrast to relatively negligible $\Delta\l=1$ correlations. 
As shown in Fig. \ref{f3}, the coefficient $K(l,\Delta\l=2)$ of increasing $\l$ is getting closer to $\overline D_{\l}/\overline C_{\l}$, while $K(l,\Delta\l=1)\rightarrow 0$. It might be also noticed that the overall magnitude of $K(l,\Delta\l=2)$ is slightly lower than $\overline D_{\l}/\overline C_{\l}$. It is attributed to approximations we made in derivation of Eq. \ref{eq15}. 
Thus, in terms of the cross-correlation coefficient our PMF-generator reproduces the statistical properties of the CMB PMF signal pretty well except for the low multipoles, where the sampling variance is large.

\subsection{The correlation of phases of CMB PMF}
The second statistic we would like to propose is based on measuring the $\Delta l=2$ phase correlation. The basic idea is to
introduce some special functions of phases, which get the minimum contribution from uncorrelated Gaussian tail and the maximum contribution from non-Gaussian tail of the phases. For this purpose the simplest trigonometric moments statistics (Fisher,1993) seem to be very useful (see Naselsky et al, 2005). Let's define the following trigonometric moments:
\begin{eqnarray}
\Cs(\l,\Delta_\l)=\frac{1}{\sqrt{\ell}}\sum_{m=1}^{\ell} \cos(\phi_{\ell,m}-\phi_{\ell+\Delta_\l,m}),\nonumber\\
\Si(\l,\Delta_\l)=\frac{1}{\sqrt{\ell}}\sum_{m=1}^{\ell} \sin(\phi_{\ell,m}-\phi_{\ell+\Delta_\l,m}),\nonumber\\
\label{eq18}
\end{eqnarray}
where $\phi_{\l,m}$ denotes the phase of $a_{\lm}$.
However, because of finite number of $m$ modes, $\Delta \l=2$ cross-correlations may exist spontaneously especially in the low multipole range. Order-of-magnitude estimate on this effect is as follows. For pure Gaussian signal of uncorrelated phases $\phi_{l,m}$
one can obtain 
\begin{equation}
\frac{1}{\sqrt{\l}}\sum_{m=1}^{{\ell}}\cos\phi_{\l,m}\sim 1 ;\hspace{0.5cm}\frac{1}{\sqrt{\l}}\sum_{m=1}^{{\ell}}\sin\phi_{\l,m}\sim 1
\label{eq19} 
\end{equation}
\begin{figure}
\centering\includegraphics[scale=.43]{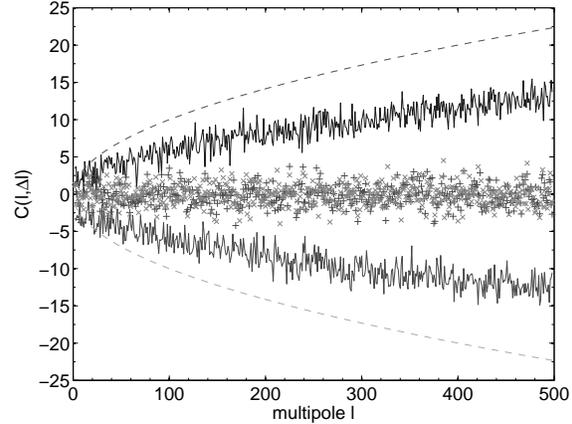}
\caption{$\Cs(\l,\Delta\l)$ for CMB PMF phases. The solid lines correspond to $\Cs(\l,\Delta\l=2)$
with $n=-5,\,\alpha=1.66$ (the upper curve) and $\alpha=-1.66$ (the lower curve). The cross and plus signs correspond to  $\Cs(\l,\Delta\l=1)$ with $\alpha=1.66$ and $\alpha=-1.66$. Two dashed lines represent the limit when $\phi_{\l+\Delta \l,m}=\phi_{\lm}$ (the top line), and $\phi_{\l+\Delta_\l,m}=\phi_{\lm}+\pi$(the bottom line).}
\label{f4} 
\end{figure}
On the other hand, $\Cs(\l,\Delta l)$ estimator has an asymptotic form $\sim \l^{\frac{1}{2}}$ 
in the limit of complete phase correlation, while  $\Si(\l)\rightarrow 0$. 
We have applied $\Cs(\l,\Delta l)$ estimator to the CMB PMF signal we had analyzed previously with a $K(\l, \Delta \l)$ statistic.
The values of $\Cs(\l,\Delta l)$ are shown in Fig. \ref{f4}, which clearly shows that $\Delta\l=2$ correlation is significant in comparison with $\Delta\l=1$ correlation just as the $K(\l, \Delta \l)$ analysis. As it is seen from non-zero values of $\Delta\l=1$ correlation, there exists spontaneous correlation due to sample variance, which is biggest at the lowest multipole. 

\section{Modification of the PMF estimator}
\subsection{flipping signs of correlations}
\begin{figure}
\centering\includegraphics[scale=.27]{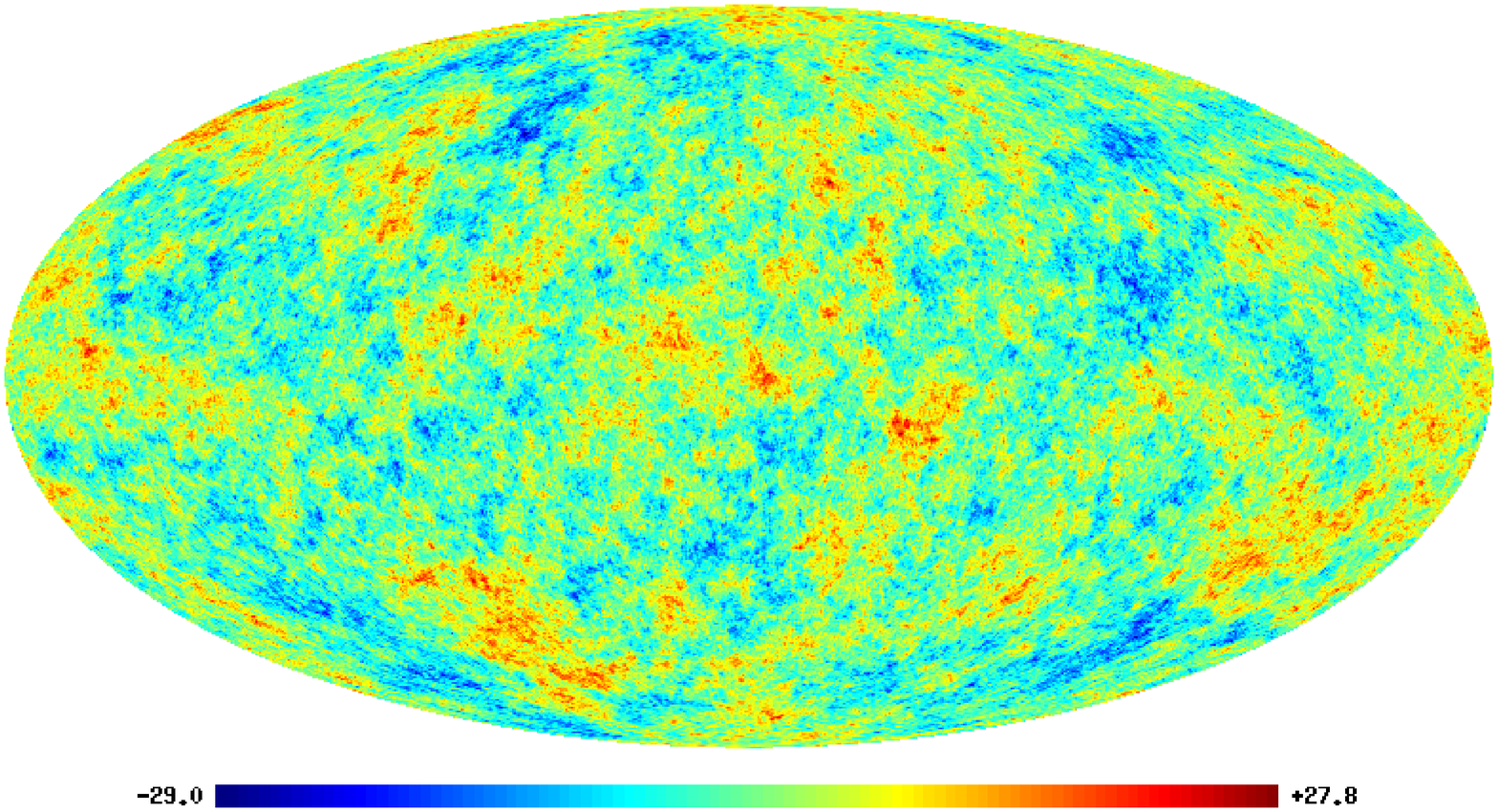}
\centering\includegraphics[scale=.27]{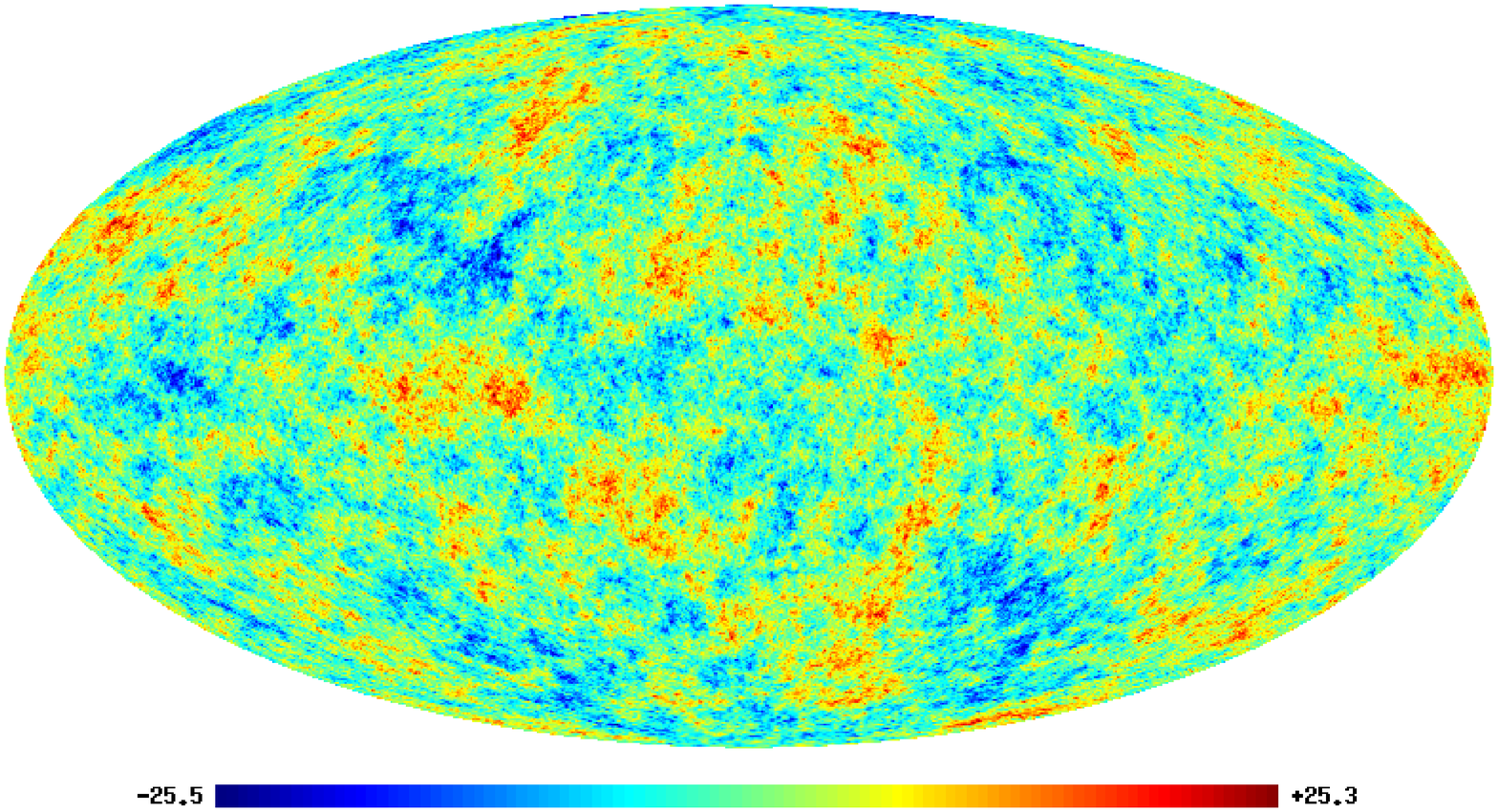}
\caption{The PMF maps obtained with $\alpha(\l)=\alpha_0 (-1)^{\l}$, 
$\alpha_0=1.66$ (top) and $\alpha_0=-1.66$ (bottom).}
\label{fflip} 
\end{figure}

Our CMB PMF generator by Eq.(\ref{eq10}) is designed in such a way that the parameter $\alpha$ is a constant of a fixed sign.  
However, $\alpha$ parameters may not be a constant of a fixed sign.
For instance, the sign of $\alpha$ parameter in Eq.(\ref{eq10}) may be a function of $\l$, depending on the PMF model.
The simplest way to extend Eq.(\ref{eq10}) is to assume that $\alpha(\l)=\alpha_0 (-1)^{\l}$. In these models, the sign of $\Delta \l=2$ correlation is alternating through multipoles.
One can easily see that the signs do not need to follow the regular (deterministic) rules such as the one above, but rather  stochastic ones.  We show the CMB PMF generated under deterministic rules in Fig. \ref{fflip}, where we may see how the morphology of the maps can change.
For even multipole $\l$, $K(\l,\Delta\l=2)$ of these CMB PMF signals correspond to the upper solid line in Fig. \ref{f3}, while 
for odd multipole $\l$, it corresponds to the lower solid in Fig. \ref{f3}.
One can see from Fig. \ref{fflip} that $\alpha(\l)$ of alternating sign leads to more homogeneous morphology of the maps, while preserving the $\Delta l=2$ correlations.

\subsection{``Brute force'' magnetism} 
As shown in Eq. \ref{eq8}, $\overline D_\l$ is comparable to $\overline C_\l$ when $n>-1$.
Let's consider the generation of the CMB PMF for the model where the relative significance of $\Delta \l=2$ is highest (i.e. $\gamma\simeq 1$).
On the other hand, for $\gamma<2$, Eq. \ref{eq16} fails to yield a real-valued $\alpha$. Extending $\alpha$ to be a complex number requires Eq. \ref{alpha_equation} to be rewritten as follows:
\begin{eqnarray}
1-\gamma \,\mathrm{Re}[\alpha]+|\alpha|^2=0. \label{alpha_equation2}
\end{eqnarray}
However, it is not difficult to show that Eq. \ref{alpha_equation2} is not satisfied by $\alpha$ of any complex value either, if $\gamma<2$.
Hence, the CMB PMF generator shown in Eq. \ref{eq10} is not adequate for $n>-1$, and we propose the following modified generator for $n>-1$:
\begin{eqnarray} 
c_{\lm}=\pm\frac{|a_{\lm}|}{|c_{\l-2,m}|}c_{\l-2,m},\label{eq20}
\end{eqnarray}
where $a_{\lm}$ is drawn from Gaussian distribution of variance $\overline C_l$ and plus (minus) sign in front of the right hand side corresponds to positive (negative) $\Delta \l=2$ correlations respectively.
\begin{figure}
\centering
\includegraphics[scale=.27]{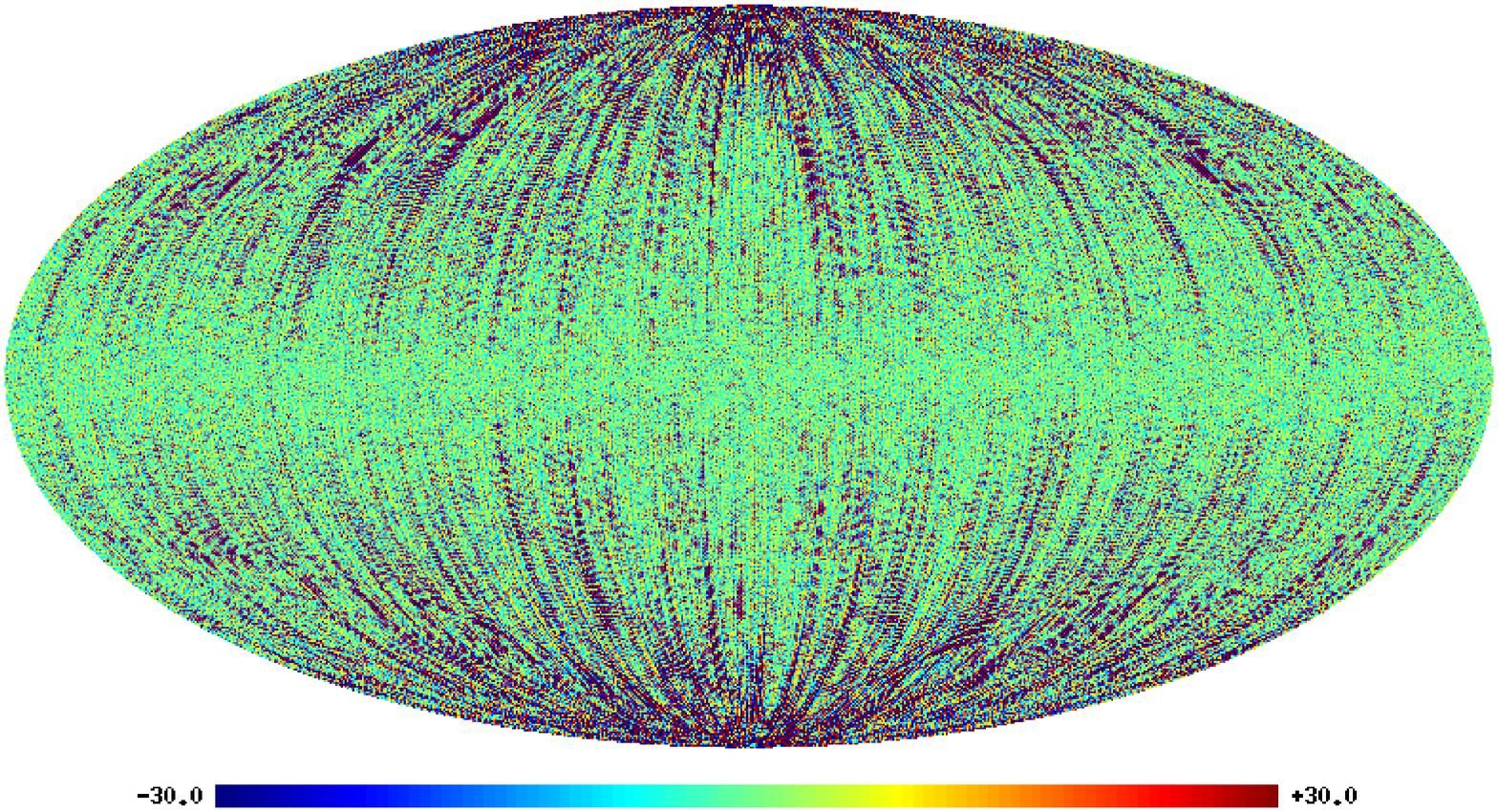}
\includegraphics[scale=.27]{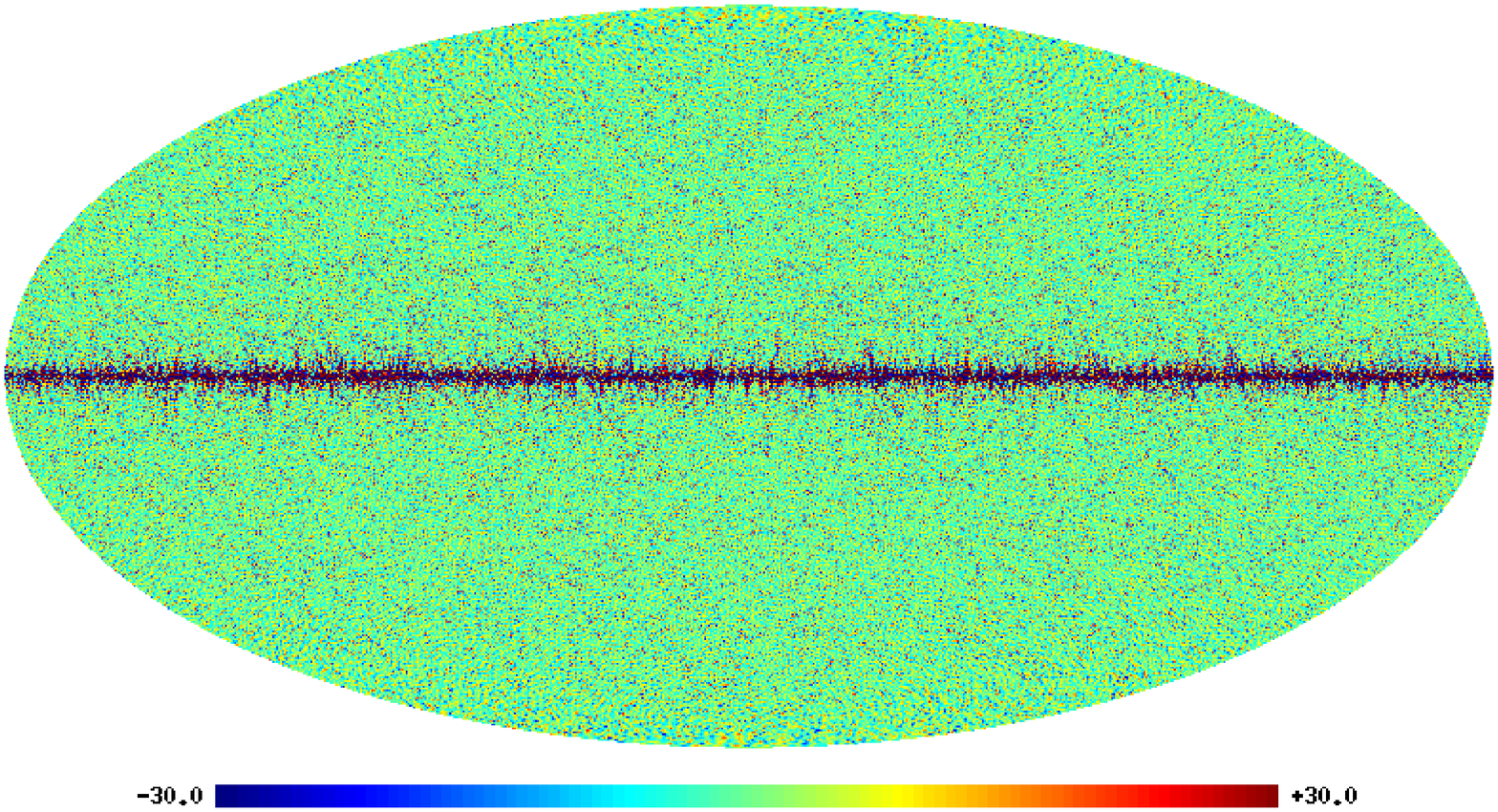}
\caption{The map for positive $\Delta \l=2$ correlations (top) and negative correlation (bottom), obtained with Eq. \ref{eq20}.}
\label{f6} 
\end{figure}
In Fig. \ref{f6} we show the CMB PMF maps generated by Eq. \ref{eq20}. The top map corresponds to positive $\Delta l=2$ correlations and the bottom map corresponds to negative ones. 
As one can see from Fig. \ref{f6}, the top map of positive $\Delta l=2$ correlations resembles a set of longitudinal stripes and have resemblance to the properties of $1/f$-noise widely discussed in the literature. 
The map of negative  $\Delta l=2$ correlations  resembles residual foregrounds, localized at the Galactic plane. 
Thus, our ``brute force magnetism'' method may be also used for modelling $1/f$ noise or Galactic foregrounds.

Since this ``brute force'' technique impose full phase coherence over $\Delta l=2$ multipoles,
the trigonometric moments of this ``brute force'' map correspond to the dash lines in Fig. \ref{f4}. 
From Eq. \ref{eq20}, we may see that phases of CMB PMF satisfy the followings:
\begin{eqnarray} 
\phi_{\l=2n,m}=\phi_{\l_0,m}+n\pi,\;\;l_0=l-2\lfloor \frac{l-|m|}{2}\rfloor,
\end{eqnarray}
where $n$ is an arbitrary integer and $\lfloor\,\rfloor$ denotes the smallest integer larger than the argument. 

It should be noted that in the ``brute force magnetism" the phases of the quadrupole component determine all the phases of even multipoles of $|m|\le 2$, while the phases of the octupole determine the phases of all odd multipoles of $|m|\le 3$. 

\section{Conclusion}
We have presented the fast generation method of the CMB PMF, which is based on the non-linear transformation of the Monte Carlo simulated Gaussian map. Our method is computationally fast and efficient, and possesses good scalability with increase in angular resolution of simulated maps. The generated maps are non-Gaussian and satisfy the $\Delta \l=2$ correlations of the CMB anisotropy generated by Alfv\'en turbulence. 

We have tested the statistical properties of our CMB PMF maps by estimating the cross-correlation and circular phase moments. As shown in this paper, both of these statistics have proved their effectiveness of our generation method. It turns out that our original generation method fails for Alfv\'en turbulence of power index $n>-1$, where
$\overline D_\l\sim\overline C_\l$.
For such Alfv\'en turbulence, we have developed a so-called ``brute force magnetism'' method, whose generated maps possess full phase coherence over $\Delta \l=2$ multipoles.
Our method can be easily extended to polarized signal and incomplete sky coverage (see Chiang and Naselsky, 2007), whose discussion will follow in separate publications.
We believe our method is quite useful for generation of non-Gaussian maps associated with the primordial magnetic field and 
it will be a valuable tool for the non-Gaussianity study of the CMB in the framework of the future PLANCK data analysis. 

\section{Acknowledgment}
 We are grateful to F. Buchet, T. Kahniashvili, S. Mattarese, E. Martinez-Gonzales and B. Wandelt for helpful discussions. 
We acknowledge the use of \healpix \footnote{\tt http://www.eso.org/science/healpix/}
package \citep{healpix} and the \glesp package \citep{glesp} for data analysis and the
whole-sky figures. This work was supported by FNU grant 272-06-0417, 272-07-0528 and 21-04-0355.

\label{lastpage}
\end{document}